\documentstyle[tighten,preprint,eqsecnum,aps]{revtex}
\begin{document}
%\draft
\preprint{IHES/P/95/99}
\title{ Testing for gravitationally preferred directions \\
 using the lunar orbit}
\author{Thibault Damour}
\address{Institut des Hautes Etudes Scientifiques, 91440
 Bures sur Yvette, France \\
 and D\'epartement d'Astrophysique Relativiste et de
 Cosmologie, Observatoire de Paris, \\
 Centre National de la Recherche Scientifique, 92195
 Meudon CEDEX, France}
\author{David Vokrouhlick\'y\cite{byline1}}
\address{Observatoire de la C\^ote d'Azur, D\'epartement
 CERGA, Avenue N. Copernic, \\ 06130 Grasse, France}
\date{\today}
\maketitle
\begin{abstract}
As gravity is a long-range force, it is {\it a priori} conceivable
that the Universe's global matter distribution select a preferred rest
frame for local gravitational physics. At the post-Newtonian
approximation, the phenomenology of preferred-frame effects is
described by two parameters, $\alpha_1$ and $\alpha_2$, the second of
which is already very tightly constrained. Confirming previous
suggestions, we show through a detailed Hill-Brown type calculation
of a perturbed lunar orbit that lunar laser ranging data have the
potential of constraining $\alpha_1$ at the $10^{-4}$ level. It is
found that certain retrograde planar orbits exhibit a resonant
sensitivity to external perturbations linked to a fixed direction in
space. The lunar orbit being quite far from such a resonance
exhibits no significant enhancement due to solar tides. Our Hill-Brown
analysis is extended to the perturbation linked to a possible
differential acceleration toward the galactic center. It is, however,
argued that there are strong {\it a priori} theoretical constraints
on the conceivable magnitude of such an effect.
\end{abstract}

\narrowtext
\section{ Introduction}
It has been recognized since many years that the lunar motion provides
a superb testing ground for relativistic gravity
\cite{dS16,B58,Ba67,N68,N73,BI85}. In particular, the Lunar Laser
Ranging (LLR) experiment has allowed one to get a very high
precision test of the equivalence principle, as well as a $1$~\% test
of the einsteinian spin-orbit coupling \cite{Sci,WND95}.
However, it has been recently pointed out that the lowest-order
perturbation analyses that have been commonly used
\cite{N68,N73,W81} to derive theoretical estimates of (null or
non-null) relativistic effects are insufficiently accurate in
view of the importance of solar tidal effects \cite{N95}. Motivated
by the results of Ref.~\cite{N95}, we presented in
Ref.~\cite{DV2} a full-fledged Hill-Brown theory of the
lunar orbit perturbation due to an hypothetical violation of the
equivalence principle. We found that the interaction with the
quadrupolar tide amplified the results of lowest-order perturbation
analyses by a very significant factor: $60$~\% increase of the naive
first-order calculation, or $40$~\% increase of the improved
first-order calculations allowing for perigee motion.
Such results raise the
question of whether similar amplification factors affect other (null
or non-null) relativistic effects in the lunar motion. To address
this question it is convenient to use the parametrized
post-Newtonian (PPN) framework (see e.g. \cite{W81}) in which
possible deviations from general relativity in the weak-field regime
are described by some parameters, $\beta -1$, $\gamma -1$,
$\alpha_1$, $\alpha_2$, etc., which vanish in Einstein's theory.

In the case of the effects linked to the Eddington post-Newtonian
parameters $\beta$ and $\gamma$ Ref.~\cite{BI85}
has indeed shown that tidal effects are numerically important.
An observationally oriented discussion of the influence of the tidal
deformation on the main effects linked to $\beta$ and $\gamma$ is
contained in Ref.~\cite{N95}, while, as we said above, the tidal
amplification of equivalence-principle-violation effects was
discussed in Refs.~\cite{N95} and, in more detail, in \cite{DV2}.

In the present paper, we study the influence of the tidal
deformation of the lunar orbit on the preferred-frame effects linked
to the parametrized post-Newtonian parameter $\alpha_1$.
We shall also discuss the effect of an hypothetical
violation of the equivalence principle of galactic origin (noting,
however, that there are strong a priori theoretical constraints
on the magnitude of such a violation).

\section{ Preferred frame effects}
As gravity is a long-range force, one might {\it a priori} expect
the Universe's global matter distribution to select a preferred rest
frame for local gravitational physics. As shown in
\cite{WN72,NW72,W81}, all preferred frame effects in the first
post-Newtonian limit are phenomenologically describable by only two
parameters, $\alpha_1$ and $\alpha_2$. These parameters are
associated with the following terms in the Lagrangian describing the
gravitational dynamics of $N$-body systems ($A,B = 1, \ldots, N$)
\widetext
\begin{mathletters}
 \label{2one}
\begin{eqnarray}
 L_{\alpha_1} &=& -{\alpha_1\over 4} \sum_{A\neq B} {Gm_A m_B\over
 r_{AB} c^2} \left({\bf v}_A^0.{\bf v}_B^0\right)\; ,
 \label{2onea} \\
 L_{\alpha_2} &=& {\alpha_2\over 4} \sum_{A\neq B} {Gm_A m_B\over
 r_{AB} c^2} \left[\left({\bf v}_A^0.{\bf v}_B^0\right)-
 \left({\bf n}_{AB}.{\bf v}_A^0\right) \left({\bf n}_{AB}.{\bf
 v}_B^0\right)\right]\; .
 \label{2oneb}
\end{eqnarray}
\end{mathletters}
\narrowtext\noindent
Here, ${\bf v}_A^0$ represents the velocity of a given body with
respect to the gravitationally preferred frame and ${\bf n}_{AB} =
({\bf r}_A - {\bf r}_B)/r_{AB}$. Many (though not all)
of the observable effects linked to $\alpha_1$ and $\alpha_2$ depend
on the choice of the gravitationally preferred frame. We shall follow
the standard assumption \cite{W81} that the latter frame, being
of cosmological origin, can be (at least approximately) identified
with the rest frame of the cosmic microwave background. This means
that the center of mass of the solar system has the velocity
${\bf w}$ with respect to the preferred frame of rest, with
$|{\bf w}| \simeq 370 \pm 10 \; {\rm km/s}$ in the direction
$(\alpha,\delta) = (168^0,-7^0)$ \cite{Pee93}.

It has been shown in Ref.~\cite{N87} that the close alignment of the
Sun's spin axis with the solar system angular momentum yields
an extremely tight bound on $\alpha_2$: $|\alpha_2| \leq
3.9\times 10^{-7}$ ($90$ \% C.L.). This limit on $\alpha_2$ is
much stronger than the existing limits on the other
post-Newtonian parameters $\beta$, $\gamma$ and $\alpha_1$.
We shall therefore neglect all $\alpha_2$ effects in this
work. Concerning $\alpha_1$, combined orbital data on the
planetary system yield \cite{H84}
\begin{equation}
 \alpha_1 = (2.1 \pm 3.1) \times 10^{-4} \qquad (90\, \%\, {\rm C.L.})
 \; , \label{2two}
\end{equation}
while binary pulsar data yield comparable or better limits
\cite{DEF92}. More precisely, PSR 1855 + 09 data yield $|\alpha_1| <
5.0 \times 10^{-4} \ (90\, \%\, {\rm C.L.})$ \cite{DEF92}, while a
recent analysis of PSR J2317 + 1439 data \cite{BC95} yield
\begin{equation}
 |\alpha_1| < 1.7\times 10^{-4} \qquad (90\, \%\, {\rm C.L.})
 \; . \label{2three}
\end{equation}
The fact that the observational limits on the $\alpha_1$ parameter
are only a factor ten better than the present limits on the (more
conservative) Eddington post-Newtonian parameters $\beta -1$ and
$\gamma -1$ stimulated recently Damour and Esposito-Far\`ese
\cite{DEF94} to propose several experiments for improving them.
Concerning their proposal to use artificial satellite motions, it has
been recognized that the currently best laser tracked satellite LAGEOS
cannot yield a better constraint on $\alpha_1$ because of badly
modeled non-gravitational forces \cite{MV1}. Another possibility
mentioned in Ref.~\cite{DEF94} (and first pointed out in \cite{N73})
concerns the lunar motion and suggests that LLR data might yield an
interesting new limit of the $\alpha_1$ parameter. However,
Refs.~\cite{N73} and \cite{DEF94} used only first-order perturbation
theory to estimate the $\alpha_1$-effects in the lunar motion. In
view of these facts (and the experience of the strong coupling with
the solar tides mentioned in Sec.~I), we decided to reassess the
quantitative value of lunar data for constraining $\alpha_1$ by
building an accurate Hill-Brown theory of the preferred-frame
perturbations of the lunar orbit.

We thus consider the 3-body Earth-Moon-Sun system (keeping the
notation of Ref.~\cite{DV2}, in particular we use the labels
$1={\rm Moon}$, $2={\rm Earth}$, $3={\rm Sun}$). Generally all these
three bodies contribute to the sum in Eq.~(\ref{2onea}), however, we
shall restrict ourselves to the ``direct'' preferred frame effects
with the subscripts $A$ and $B$ spanning only $1$ (Moon) and $2$
(Earth). It is easy (though not trivial) to verify that the ``tidal''
preferred frame effects, involving the subscript $3$ (Sun) in
Eq.~(\ref{2onea}), are several orders of magnitude smaller than the
direct effects. Because of their observational irrelevance, we also
omit from our discussion several terms in Eq.~(\ref{2onea}) which are
equivalent to a nearly constant redefinition of the locally measured
gravitational constant. The dominant preferred frame effects are then
contained in the following three contributions to the lagrangian (we
factorized the Earth-Moon reduced mass $\mu_{12} \equiv m_1 m_2/m_0$,
$m_0 = m_1+m_2$, from the lagrangian)
\begin{mathletters}
 \label{2four}
\begin{eqnarray}
 R^{(1)}_{\alpha_1} &=&  -{\alpha_1\over 2c^2} {Gm_0 \over r} X_{21}
  \left({\bf w}.{\bf v}\right)\; , \label{2foura} \\
 R^{(2)}_{\alpha_1} &=&  -{\alpha_1\over 2c^2} {Gm_0 \over r} X_{21}
  \left({\bf v}_0.{\bf v}\right)\; , \label{2fourb} \\
 R^{(3)}_{\alpha_1} &=&  -{\alpha_1\over c^2} {Gm_0 \over r}
  \left({\bf w}.{\bf v}_0\right)\; . \label{2fourc}
\end{eqnarray}
\end{mathletters}
Here, ${\bf r}\equiv {\bf x}_1 - {\bf x}_2$ is the geocentric
lunar position vector and ${\bf v}\equiv d{\bf r}/dt$ its
velocity, ${\bf v}_0$ is the velocity of the Earth-Moon
center of mass motion around the Sun, and $X_{21}\equiv X_2-X_1$,
with the mass ratios $X_1\equiv m_1/m_0$ and $X_2\equiv m_2/m_0
\equiv 1-X_1$. In the following section, we treat successively the
perturbations of the lunar orbit associated with the three terms
(\ref{2foura})--(\ref{2fourc}).

\section{ Hill-Brown treatment of preferred frame effects}
\subsection{ The method in brief}
Since the Hill-Brown approach to lunar motion represents a
classic tool of celestial mechanics treated with care in the
literature (e.g. \cite{Brown,A26,BC,GS86}) we outline its concept
only very briefly, focusing mainly on the particularities of the
method involved in the present study. We also refer the interested
reader to Ref.~\cite{DV2} for more details and used notation.

Following Hill we start by considering the planar
Earth-Moon-Sun $3$-body problem with the Earth-Moon center-of-mass
on a circular orbit around the Sun (the so called ``Main Problem'').
The near circular lunar motion is investigated in an Earth-centered
coordinate system $(X,Y)$ rotating with the angular velocity $n'$
corresponding to the solar motion around the Earth-Moon
center-of-mass (the Sun thus rests on the axis $X$). Apart from the
Earth direct gravitational action, the quadrupolar piece of the solar
(tidal) gravitational potential is also taken into account. The
reduced lagrangian of the lunar motion then reads
\begin{equation}
 L_{\rm Hill} = {1\over 2}\left({\dot X}^2+{\dot Y}^2\right)+
 n'\left(X{\dot Y}-Y{\dot X}\right) + R_{\rm Hill}\; ,
 \label{3one}
\end{equation}
with
\begin{equation}
 R_{\rm Hill} = {Gm_0\over \sqrt{X^2+Y^2}} + {3\over 2}n'^2 X^2
 \label{3two}
\end{equation}
(the overdot means $d/dt$).
Although simplified, the theory entails the most important
part of the solar tidal deformation of the lunar motion. Instead of
the usual keplerian ellipse Hill chooses for the intermediary
lunar orbit a periodic solution (with particular symmetries) of
(\ref{3one}), the so called ``variational curve''. In the following,
we shall investigate the forced perturbations of the variational orbit
due to the additional lagrangian terms (\ref{2four})\footnote{
We recall that the variational orbit is not
a general solution of the system (\ref{3one}). Apart from the
``forced'' perturbations related to a new physical cause it admits
also ``free'' perturbations covering the classical notion of the lunar
orbit eccentricity and its perigee drift due to the solar action.
For simplicity, we omit in this study a natural coupling of the
two types of perturbations neglecting thus the (small) lunar
eccentricity (and inclination) corrections to the preferred frame
perturbations.}.

A very convenient parametrization of the Hill problem consists of
replacing $(X,Y)$ by the complex conjugated quantities $(w,{\bar w})$
defined by ($i$ is the complex unit)
\begin{mathletters}
 \label{3three}
\begin{eqnarray}
 X+ i Y &=& {\tilde a}\zeta (1+w) \; , \label{3threea} \\
 X- i Y &=& {\tilde a}\zeta^{-1} (1+{\bar w}) \; , \label{3threeb}
  \\ \zeta &=& e^{i \tau} \; , \label{3threed} \\
 \tau &=& (n-n')t + \tau_0 \; , \label{3threec}
\end{eqnarray}
\end{mathletters}
where $n$ denotes the mean lunar motion around the Earth and where the
$\tau$ variable represents the mean geocentric
angular separation of the Moon and the Sun. Following
Ref.~\cite{BI85}, the fiducial lunar semi-major axis ${\tilde a}$ is
defined by
\begin{equation}
 {G m_0 \over (n-n')^2 {\tilde a}^3} = \kappa(m)
 \; , \label{3four}
\end{equation}
where
\begin{equation}
 \kappa(m) = 1+2 m+{3\over 2} m^2\; . \label{3five}
\end{equation}
and $m\equiv n'/(n-n')$ is Hill's expansion parameter. (In the
actual case of the Moon, $m_{\rm Moon} =0.0808489375\ldots$.) The
Lagrange dynamical equations of the variational motion then read
\begin{equation}
 L(w,{\bar w}) =  W_{\rm Hill}(w,{\bar w})\; , \label{3six}
\end{equation}
where we denoted
\begin{equation}
 L(w,{\bar w}) = D^2w+ 2(1+ m) Dw + {3\over 2} \kappa(m)
 (w+{\bar w})\; , \label{3seven}
\end{equation}
with $D \equiv d/(id\tau) = \zeta d/d\zeta$, and the Hill source
terms
\begin{equation}
 W_{\rm Hill}(w,{\bar w}) = -{3\over 2} m^2 \zeta^{-2} (1+{\bar
 w})+ \kappa(m) Q(w,{\bar w}) \; . \label{3eight}
\end{equation}
The nonlinear source function $Q(w,{\bar w})$ and its development in
terms of $(w,{\bar w})$ can be found for instance in Ref.~\cite{DV2}.

When considering an extra perturbation of the lunar motion, such
as (\ref{2four}) in the case of preferred frame effects, we have
to include an additional source function on the right hand side of
Eq.~(\ref{3six}) given by
\begin{equation}
 S\left(w,{\bar w};Dw,D{\bar w}\right) = D{\partial \over \partial
  D{\bar w}} G - {\partial \over \partial{\bar w}}
  G \; , \label{3nine}
\end{equation}
in terms of the ``generating function'' $G(w,{\bar w};Dw,D{\bar w})
\equiv 2(m/n'{\tilde a})^2 R$.

Our method of solution of the system (\ref{3six}) to (\ref{3nine})
consists of consecutive iterations, where at each stage one constructs
a particular right-hand side source based on the results of the
previous iterations. Details can be found in Ref.~\cite{BI85} or
Appendix~B of Ref.~\cite{DV2}. Let us only
point out that, contrary to the simpler case of the synodic
lunar perturbations due to an hypothetic violation of the
equivalence principle studied in Ref.~\cite{DV2}, the generic form
of the right hand side source term now reads
\begin{equation}
 W_\star(\alpha) = W_{-\alpha} \zeta^{-\alpha} + W_{\alpha}
 \zeta^{\alpha} \; , \label{3ten}
\end{equation}
where we allow for: (i) complex functions $W_{-\alpha}$ and
$W_{\alpha}$, and (ii) any {\it real} (non integer) values of the
powers $\alpha$. Inversion of the linear problem
$L(w_\star(\alpha),{\bar w}_\star(\alpha)) = W_\star(\alpha)$
($\alpha\neq 0$) has a simple form $w_\star(\alpha) \equiv
w_{-\alpha}\zeta^{-\alpha} + w_{\alpha} \zeta^{\alpha}$ with
\widetext
\begin{mathletters}
 \label{2eleven}
\begin{eqnarray}
 w_\alpha &= & {1\over \Delta_\alpha(m)}\left\{\left[\alpha^2
 -2\left(1+m\right)\alpha +{3\over 2} \kappa\right]
 W_\alpha - {3\over 2} \kappa {\overline{W}_{-\alpha}}\right\} \; ,
 \label{3elevena} \\
 w_{-\alpha} &= & {1\over \Delta_\alpha(m)}\left\{\left[\alpha^2
 +2\left(1+m\right)\alpha +{3\over 2} \kappa\right]
 W_{-\alpha}- {3\over 2} \kappa {\overline{W}_{\alpha}}\right\} \; ,
 \label{3elevenb}
\end{eqnarray}
\end{mathletters}
\narrowtext
\noindent and
\begin{equation}
 \Delta_\alpha(m) \equiv \alpha^2\left[\alpha^2+3 \kappa-
  4\left(1+m\right)^2\right] \label{3twelve}
\end{equation}
[The solution corresponding to $\alpha=0$ is identical with that given
in Eq.~(2.52a) of Ref.~\cite{DV2}.] Some values of the power $\alpha$
in (\ref{3twelve}) may lead to a significant amplification of the
effect due to the smallness of the corresponding
denominator $\Delta_\alpha(m)$. Of particular interest for our
present work is the case where $\alpha = 1+m$ which yields the small
denominator
\begin{equation}
 \Delta_{1+m}(m) \equiv {3\over 2}m^2 (1+m)^2
 \; . \label{3thirteen}
\end{equation}
In the next section we shall see that it appears in the
sidereal excitation of the lunar orbit.

Because of the background motivation of our work, related to the
LLR experiment, we are essentially interested in the perturbation
of the radial geocentric distance of the Moon given by
\begin{equation}
 r^2  = {\tilde a}^2 (1+w)(1+{\bar w})\; . \label{3fourteen}
\end{equation}
Performing a variation of this quantity, keeping only linear
terms in the perturbation, we obtain
\begin{equation}
 {\delta r \over {\tilde a}}  =  \Re e\,\left[\left(
  {1+{\bar w} \over 1+w}\right)^{1/2} \delta w\right] \; ,
 \label{3fifteen}
\end{equation}
for the searched perturbation in radial coordinate. Remembering that
$w = {\cal O} (m^2)$, to lowest order in the $m$ parameter, the radial
oscillation can be expressed by the simple formula: $\delta
r/{\tilde a} \simeq (w+{\bar w})/2$.

In the rest of this section we investigate the forced perturbations
of the lunar variational orbit related to the
three preferred-frame lagrangian terms (\ref{2four}). Finally, we
note that albeit the iteration scheme mentioned previously is
straightforward it represents a huge algebraic manipulation exercise.
We thus employed the powerful dedicated algebraic computer system
MINIMS developed by M.~Moons from the University of Namur (Belgium)
\cite{Moons} to perform this task. The lowest two orders of
the results have been, however, checked by hand computations.

\subsection{ Potential $R_{\alpha_1}^{(1)}$}
Firstly, we focus on the source term (\ref{2foura}). The corresponding
generating function $G$ reads
\widetext
\begin{equation}
 G(w,Dw) = -i {\hat \epsilon}_1 \left({{\tilde a} \over r}
 \right)\Bigl\{\left[Dw+\left(1+m\right)\left(1+w\right)\right]
 \zeta^{1+m} e^{-i\phi}
 + \left[D{\bar w}-\left(1+ m\right)\left(1+{\bar
 w}\right)\right]\zeta^{-(1+m)} e^{i\phi}\Bigr\} \;
 , \label{3sixteen}
\end{equation}
where
\begin{equation}
 {\hat \epsilon}_1 = {\alpha_1 \over 2} X_{21} {\cal C}
 {|{\bf w}| v_0\over c^2} \left({{\tilde a}\over a'}\right) {{\hat
 \kappa}(m)\over m} \; , \label{3seventeen}
\end{equation}
and where one must express $r$ in terms of $w$ through
Eq.~(\ref{3fourteen}).
Here, $a'$ is the radius of the (circular) solar orbit in the
Earth-Moon center-of-mass frame and $v_0 = n' a'$ its (circular)
velocity, ${\cal C}(\simeq 0.98)$ is the cosine of the ecliptic
latitude of the unit vector ${\bf w}^0$, and $\phi$ is
a longitude angle of ${\bf w}^0$ measured from the lunar (and
solar) position at time $t_0$ corresponding to an arbitrary new-moon
phase. For instance, if we choose the last new-moon phase in this
century, occurring at MJD51~521.2, we obtain $\phi= 267.2^0$.
Inserting this expression into
(\ref{3nine}) we obtain the source function, to be added to the right
hand side of the Hill equation (\ref{3six}), in the following form
\begin{eqnarray}
 S\left(w,{\bar w};Dw,D{\bar w}\right) &=& -i{{\hat
 \epsilon}_1\over 2}\left({{\tilde a}\over r}\right) \biggl\{Dw
 \left[\zeta^{1+m}{e^{-i\phi}\over 1+{\bar w}}-
 \zeta^{-(1+m)} {e^{i\phi}\over 1+ w}\right]
 \nonumber \\
 & & \qquad\qquad\qquad+ \left(1+m\right) \left[{1+w\over
 1+{\bar w}} \zeta^{1+m}e^{-i\phi}- \zeta^{-(
 1+m)} e^{i\phi}\right]\biggr\}\; .
 \label{3eighteen}
\end{eqnarray}
\narrowtext

Working out the iterative solution mentioned above one realizes that
this perturbation yields a wide spectrum of radial and
longitudinal oscillations of the lunar orbit (compare also with
the less accurate solution in Ref.~\cite{N73}). However, a detailed
analysis shows that only two of them are sufficiently amplified to
give an observably interesting signal: (i) terms with frequency equal
to the mean sidereal lunar motion $n$ (having a period of about
$27.^{\rm d}32$), and (ii) terms with frequency equal to $n-2n'$
(having a period of about $32.^{\rm d}13$). Both periods are evaluated
for the lunar orbit. Hereafter we discuss properties of both of them
starting with the sidereal terms.

The perturbation series giving the sidereal-frequency radial
oscillations of the lunar orbit reads
\begin{eqnarray}
 {\delta_\epsilon r\over {\tilde a}} = {2{\hat \epsilon}_1\over 3
  m^2} S^{(1)}_{\alpha_1}\left(m\right)\sin\left[n\left( t-t_0\right)-
  \phi\right] \; ,  \label{3nineteen}
\end{eqnarray}
with
\begin{eqnarray}
 S^{(1)}_{\alpha_1}\left(m\right) &=& 1 -{67\over 8}m +{395\over
  8}m^2 - {103007\over 384}m^3 \nonumber \\
 & & + {3327349\over 2304}m^4 + {\cal O}(m^5) \; . \label{3twenty}
\end{eqnarray}
Table~\ref{tab1} gives the coefficients of the series $S^{(1)
}_{\alpha_1}(m)$ up to the ninth order. The second
column of the table indicates the numerical contribution of the
corresponding term to the total value of the series for the lunar
orbit, i.e. $m=m_{\rm Moon} = 0.0808489375\ldots$ \cite{GS86}.
Two important features are to be noticed: (i) a significant
contribution of the higher
order corrections to the total value of the series $S^{(1)
}_{\alpha_1}(m_{\rm Moon})$, and (ii) a geometric-like character
of this series clearly pronounced after a few terms. The
second property suggests the presence of a pole near the
value\footnote{Such a value corresponds to a retrograde orbit with
a sidereal period (for Earth satellites) $T=2\pi /|n| =82.4$ days.}
$m_{cr} \simeq -0.18407$. Taking a Pad\'e approximant of the series
$S^{(1)}_{\alpha_1}(m)$ confirms the presence of a unique root of the
denominator located at the value $m_{cr} = -0.18407$ with an error of
about $10^{-5}$.

The physical origin of this pole can be easily understood by using
the following argument. When solving the problem by traditional
first-order perturbation techniques (see e.g. \cite{N73,DEF94}),
one finds that the sidereal orbit oscillation induced
by an external force linked to a fixed direction in space contains,
in the denominator, the secular rate of the perigee longitude ${\dot
\varpi}$. This agrees with the intuitive idea that a
spatially ``frozen'' orbit (not moving its pericenter in fixed
space) is ``resonantly sensitive'' to constant forces. As a check of
this idea, we have multiplied the $S^{(1)}_{\alpha_1}(m)$ series by
${\dot \varpi}(m) = n \left[{3\over 4} m^2 + {177 \over 32}
m^3 +\cdots \right]$ as given by Andoyer up to order $m^9$ in
Ref.~\cite{A26}. The resulting series, say
$\overline{S}_{\alpha_1}^{(1)} (m) = 4 S_{\alpha_1}^{(1)}(m)
{\dot \varpi}(m)/(3 nm^2) =1-m+{313 \over 64} m^2+\cdots$,
is much more tame, showing that the main characteristics of
$S^{(1)}_{\alpha_1}(m)$ are entailed in the factor $[{\dot
\varpi}(m)]^{-1}$. The difference of our more precise solution
(\ref{3nineteen}) with the previous ones \cite{N73,DEF94} (when the
latter are improved by using the full value of the perigee advance)
is essentially contained in the value of the residual series
$\overline{S}_{\alpha_1}^{(1)} (m)$. We find that, in the lunar case,
$\overline{S}_{\alpha_1}^{(1)} (m_{\rm Moon}) \simeq 0.956$. In
contrast with the case of equivalence-principle-violation effects,
we see therefore that preferred-frame effects exhibit no significant
enhancement genuinely linked to the tidal deformation of the lunar
orbit. This results holds for the other effects discussed below and
is basically attributable to the fact that the actual lunar orbit is
quite different from the ``spatially frozen'' resonant orbit (while
it is rather near the orbit resonant for solar-directed
equivalence-principle-violation effects).

The principal quantitative information of the previous analysis is
given by the amplitude of the sidereal oscillation of the lunar orbit
(\ref{3nineteen})\footnote{ Beware, however, that for the final
determination of the $\alpha_1$ constraint through LLR data
analysis, the particular phase $\phi$ is very important.
Moreover, we shall see that the preferred frame perturbations of
the lunar orbit act also with several other frequencies and it is
their combined influence which determines the full effect to be
searched for in the LLR data.}. Employing the definition (\ref{3four})
of the auxiliary lunar semimajor axis ${\tilde a}$ [and a Pad\'e
approximant value of $S_{\alpha_1}^{(1)} (m) : S_{\alpha_1}^{(1)}
(m_{\rm Moon}) \simeq 0.5465$] we have
\widetext
\begin{equation}
 |\delta_\epsilon r| \simeq {\alpha_1\over 3} X_{21} {\cal C}
  {|{\bf w}| v_0\over c^2} \left[{ \kappa \chi^2 \over m^5}
  \right]^{1/3} S^{(1)}_{\alpha_1}(m)\, a'\simeq 4780
  \times\,\alpha_1 {\cal C}\;\; {\rm [cm]}\; , \label{3twentyone}
\end{equation}
\narrowtext \noindent
where $\chi\equiv (1+m_3/m_0)^{-1}$. The current published {\it
accuracy} of the lunar ranging measurements performed by the
CERGA team is 14 millimeters. Recent technical improvements are
giving a timing {\it precision} of about 6 millimeters (C.~Veillet,
private communication). If the latter precision level can be turned
into an accuracy level, the result (\ref{3twentyone}) suggests that
the LLR data should soon be able to constrain $\alpha_1$ at the
$1\times 10^{-4}$ level or better (given the phase information and the
presence of several $\alpha_1$-effects at different frequencies).

The value $m=0$ of the Hill
parameter is apparently another singularity of our ``ranging
formula'' (\ref{3nineteen}). However, because we neglected the
Earth quadrupole and the other higher multipoles of the Earth gravity
field, we cannot extend our solution to near-Earth satellite
orbits ($m \simeq 0$). This regime has been thoroughly discussed
in Ref.~\cite{DEF94}.
In a first approximation we can, however, match smoothly
the solution of Ref.~\cite{DEF94}, accounting basically for
the Earth quadrupole, and the solution presented in the present
study, accounting in detail for the third-body (Sun)
perturbations, by adding the Earth quadrupole contribution
to ${\dot \varpi}(m)$ after having factorized it as denominator of the
series $S^{(1)}_{\alpha_1} (m)$. [Actually, as Andoyer's series is not
accurate enough to localize precisely the zero at $m=m_{cr}$, we
found better to add the quadrupole contribution to the denominator
of a Pad\'e approximant of $S^{(1)}_{\alpha_1}(m)$.]
Figure~\ref{fig1} shows the
synthesis of the two effects. The arrow points the singularity
$m=m_{cr}$, while the two points $M$ and $L$ stand for the Moon and
the artificial (retrograde) satellite LAGEOS, respectively (without
taking into
account the LAGEOS inclination). We can see that none of the two
bodies (an equatorial satellite at the LAGEOS altitude or the Moon)
is the best candidate for testing preferred frame effects, but
that (as mentioned in \cite{DEF94}) a high orbit artificial body with
a period of about $30$ hours optimizes the sensitivity to the
$\alpha_1$ parameter (among prograde orbits). On the other hand, one
should be aware of the fact that the motion of artificial bodies is
typically influenced by many non-gravitational forces, some of which
are difficult to be predicted and/or carefully modeled. For instance,
this is the reason why the LAGEOS satellite is currently less
suitable for constraining the $\alpha_1$ parameter than the Moon
\cite{MV1}, which is a nearly perfect ``drag free Earth-satellite''.
Therefore the lunar data stand out as a potentially important source
for the study of preferred frame effects.

The intricate interaction of the variational curve perturbations
with the underlying tidal deformation leads also to a slowly
convergent series for the perturbations at the $(n-2n')$-frequency.
The final result for the radial oscillations formula reads
\begin{equation}
 {\delta_\epsilon r\over {\tilde a}} = - {5\over 4} {{\hat
  \epsilon}_1\over m} S^{\prime(1)}_{\alpha_1}\left(m\right)\,
  \sin\left[\left(n-2n'\right) \left(t-t_0\right)+\phi\right]
  \; . \label{3twentytwo}
\end{equation}
with
\begin{equation}
 S^{\prime(1)}_{\alpha_1}\left(m\right) = 1 -{43\over 6}m +{28867
  \over 720}m^2  - {468391\over 2160}m^3 +{\cal O}(m^4)
  \; . \label{3twentythree}
\end{equation}
The coefficients of $S^{\prime(1)}_{\alpha_1}(m)$ up to the eight
order are given in Table~\ref{tab2}. The intimate coupling of this
frequency with the sidereal frequency results in the coincidence of
the pole in the $m$-series $S^{\prime(1)}_{\alpha_1}(m)$ and
$S^{(1)}_{\alpha_1}(m)$. Numerically $S_{\alpha_1}^{\prime(1)}
(m_{\rm Moon})\simeq 0.6035$, and the lunar orbit sensitivity to
the $\alpha_1$
perturbation on this frequency is given by $|\delta_\epsilon r|
\simeq 800\times \alpha_1\;\;{\rm [cm]}$, approximately $5.98$
times smaller than for the principal sidereal effect. Notice,
however, that this term in the spectrum of the preferred
frame lunar perturbations may bound the $\alpha_1$ parameter as
efficiently as the sidereal term if it turns out that there is
significantly less noise at this frequency.

\subsection{ Potential $R_{\alpha_1}^{(2)}$}
A special character of this term is due to its independence on the
choice of the gravitationally preferred frame. It would
be theoretically appealing if this term could significantly
contribute to constraining $\alpha_1$.
Unfortunately, we shall demonstrate that the significance of this
perturbation faces two obstacles: (i) its amplitude is small, and
(ii) it acts with a synodic frequency, the same as the other
phenomena tested through the LLR experiment (e.g. the classic
equivalence-principle-violation effect; \cite{W81,N95,DV2}).

In the context of the Main Lunar Problem we consider a circular solar
orbit around the Earth. The velocity ${\bf v}_0$ thus becomes $-v_0
{\bf e}_Y$ ($v_0 = n'a'$) in the rotating (Hill) coordinate system
introduced in Sec.~II. The generating function $G$ reads
\widetext
\begin{equation}
 G(w,Dw) = {\hat \epsilon}_2 \left({{\tilde a} \over r}\right)
 \Bigl\{ \left[Dw+\left(1+m\right)\left(1+w\right)\right]\zeta
  -\left[D{\bar w}-\left(1+m\right)\left(1+{\bar w}\right)\right]
 \zeta^{-1}\Bigr\} \; , \label{3twentyfour}
\end{equation}
with
\begin{equation}
 {\hat \epsilon}_2 = {\alpha_1 \over 2} X_{21} \left({ v_0\over c}
 \right)^2 {\kappa(m)\over m} {{\tilde a}\over a'} \; .
 \label{3twentyfive}
\end{equation}
Then the source function can be easily calculated by using
(\ref{3nine})
\begin{eqnarray}
 S\left(w,{\bar w};Dw,D{\bar w}\right) &=& {{\hat \epsilon}_2 \over 2}
 \left({{\tilde a}\over r}\right) \biggl\{\zeta^{-1}\left(1-m
 \right)+ \zeta\left(1+m\right){1+w\over 1+{\bar w}} \nonumber
 \\ & &\qquad\qquad\quad + Dw \left[{\zeta^{-1}
 \over 1+ w}+{ \zeta\over 1+ {\bar w}}\right]\biggr\}\; .
 \label{3twentysix}
\end{eqnarray}
\narrowtext
\noindent
The close similarity with the classical equivalence-principle-violation
effect studied in \cite{DV2} consists
of the fact that the source function (\ref{3twentysix}) excites the
odd powers of $\zeta$, resulting in (radial and longitudinal)
oscillations of the lunar orbit with the synodic frequency, aliasing
with the equivalence-principle-violation effect \cite{N95,DV2}.

We have learnt in Refs.~\cite{N95,DV2} that any synodic signal
is particularly amplified by the presence of a pole singularity
occurring for a prograde orbit about 68~\% larger than the lunar
orbit ($m_{cr} = 0.19510399\ldots$). The corresponding ranging formula
reads
\begin{equation}
 {\delta_\epsilon r\over {\tilde a}} = -{{\hat \epsilon}_2\over 2m}
  S^{(2)}_{\alpha_1}(m) \cos\tau \; , \label{3twentyseven}
\end{equation}
where $S^{(2)}_{\alpha_1}(m)$ is a series in the Hill parameter $m$,
whose numerical value for the lunar orbit is found to be $1.3022$. The
amplitude of the synodic oscillation (\ref{3twentyseven}) of the lunar
orbit thus reads $|{\delta_\epsilon r}| \simeq 57\times \alpha_1
\;\; {\rm [cm]}$, too small to compete with the much greater
sensitivity of the lunar orbit to the equivalence principle
violation term \cite{N95,DV2}.

\subsection{ Potential $R_{\alpha_1}^{(3)}$}
{}From Eq.~(\ref{2fourc}) we see that this perturbing term is
equivalent to a variation of the gravitational constant $G$ (not
to be confused with the generating functions $G(w,Dw)$) with a
period of one year (see e.g. \cite{DEF94}). The analysis of this
term involves a small denominator $\Delta_{\alpha=m} \propto
 - m^2$ which, however, cancels out in the radial
oscillation \cite{DEF94}. As found in Ref.~\cite{N73}, the remaining
signal still exhibits an interesting sensitivity to the $\alpha_1$
parameter.

The generating function $G$ equivalent to the
$R_{\alpha_1}^{(3)}$ reads
\begin{equation}
 G(w,Dw) = i{\hat \epsilon}_3 \left({{\tilde a} \over r}\right)
 \left( \zeta^m e^{-i\phi} -\zeta^{-m} e^{i
 \phi}\right) \; , \label{3twentyeight}
\end{equation}
with
\begin{equation}
 {\hat \epsilon}_3 = \alpha_1 {\cal C} {|{\bf w}| v_0\over c^2}
 \kappa(m)\; , \label{3twentynine}
\end{equation}
and the resulting source function is
\begin{equation}
 S\left(w,{\bar w}\right) =  {{\hat \epsilon}_3 \over 2}
 \left({{\tilde a}\over r}\right){i \over 1+{\bar w}}
 \left(\zeta^m e^{-i\phi} -\zeta^{-m} e^{i
 \phi}\right) \; . \label{3thirty}
\end{equation}
Using the previous scheme of solving the perturbation equations we
obtain
\begin{equation}
 {\delta_\epsilon r \over {\tilde a}} = {\hat \epsilon}_3
 S^{(3)}_{\alpha_1}(m) \sin \left[n'\left(t-t_0\right)-\phi\right]
 \label{3thirtyone}
\end{equation}
for the expected contribution to the radial perturbation of the
lunar orbit. The first terms of the series $S^{(3)}_{\alpha_1}(m)$
are listed in Table~\ref{tab3}. The magnitude of the oscillations
(\ref{3thirtyone}) is about $|\delta_\epsilon r| \simeq 4550\times
\alpha_1 {\cal C}\;\;{\rm [cm]}$, comparable to the sidereal effect
coming from $ R_{\alpha_1}^{(1)}$. The value $\kappa (m)
S_{\alpha_1}^{(3)} (m) \simeq 0.9643$ shows that tidal effects are
not very important. We learned from J.G.~Williams (private
communication) that the prospects of decorrelating the effect
(\ref{3thirtyone}) from other annual effects is high. If this is
confirmed, Eqs.~(\ref{3twentyone}) and (\ref{3thirtyone}) would
be the two best probes for constraining $\alpha_1$.

\section{ Polarization of the lunar orbit by a galactic differential
acceleration}
Besides preferred-frame effects, other perturbing forces can be
linked to some fixed direction in space. This would be, in
particular, the case if the Earth and the Moon would fall with a
different acceleration toward the center of the Galaxy. Years ago
this possibility has been mentioned in relation with a possible
violation of the equivalence principle linked to the gravitational
binding energy of planets \cite{N70}. More recently, this idea has
been revived within the context of possible strong-gravitational
field effects in neutron stars, and has led, through the use of
existing binary pulsar data, to new tests of strong-field gravity
\cite{DS91}. Still more recently, the idea surfaced again with a
different motivation: the possibility that the coupling between
ordinary (visible) matter and galactic dark matter violate the
equivalence principle \cite{S93}. The latter suggestion led to new
galactic-related laboratory tests of the equivalence principle
\cite{SAHS93,Su94}, as well as to a corresponding reanalysis
of lunar laser ranging data \cite{N94,NMS95,MSS}.
Before applying our Hill-Brown algorithm to the latter problem (with
the result that we find no unexpected amplification), we wish to
emphasize that there are strong {\it a priori} theoretical
constraints (using existing observational data) on the conceivable
magnitude of any ``dark matter effect''. These constraints diminish,
in our opinion, the theoretical significance of the results of
Refs.~\cite{SAHS93,Su94,NMS95}.

We assume a field-theoretic framework (as is always the case in
recent discussions concerning possible violations of the equivalence
principle; e.g. \cite{S93,Su94}). Within such a framework,
any effect on visible matter due to a new (non-Einsteinian)
long-range field generated by dark matter is necessarily
proportional to the product $\alpha_V \alpha_I$ of two coupling
constants: $\alpha_V$ measuring the coupling of the field to visible
matter, and $\alpha_I$ the coupling to invisible (dark) matter. We
normalize these coupling constants with respect to the usual
Einsteinian coupling so that the effective gravitational constant
between bodies $A$ and $B$ reads $G_{AB} = G_* (1 \pm \alpha_A
\alpha_B)$ where $G_*$ is a bare Newtonian constant and where the
plus (minus) sign holds for a spin 0 (spin 1) mediating field. To
fix ideas, let us consider the case of a scalar field (our argument
goes through in both cases, but is newest in the scalar case, the
vector case having been already discussed in \cite{SAHS93}, though
with a less stringent constraint on $\alpha_I$). Equivalence principle
tests probe the composition dependence of the visible couplings:
$\alpha_A^{(V)} - \alpha_B^{(V)} \not = 0$. Let us denote by
$f_{AB}$ the fractional modification of the average coupling to
visible matter $\alpha_V$ in a differential composition-dependent
experiment: $\alpha_A^{(V)} - \alpha_B^{(V)} = f_{AB} \alpha_V$. The
ordinary tests of the equivalence principle (using visible sources)
measure the fractional differential acceleration
\begin{equation}
 \eta_{AB}^{VV} \equiv \left( {\Delta a \over a} \right)_{AB}^{VV} =
 (\alpha_A^{(V)} -\alpha_B^{(V)}) \alpha_V = f_{AB} \alpha_V^2 \, .
 \label{4one}
\end{equation}

These experiments give us a handle (in practice, an upper limit) on
the magnitude of $\alpha_V : \alpha_V = (\eta_{AB}^{VV} /
f_{AB})^{1/2}$. [For simplicity, we do not put absolute value signs
around $\alpha$, $\eta$ and $f$.] Then the tests of the equivalence
principle using the same pair of visible bodies, and an invisible
source (e.g. the galactic dark matter) measure,
\begin{equation}
 \eta_{AB}^{VI} \equiv \left({\Delta a \over a}\right)_{AB}^{VI} =
 (\alpha_A^{(V)} - \alpha_B^{(V)}) \alpha_I = f_{AB} \alpha_V
 \alpha_I \, . \label{4two}
\end{equation}
Inserting the previous value of $\alpha_V$ into (\ref{4two}) yields
\begin{equation}
 \eta_{AB}^{VI} = (f_{AB})^{1/2} (\eta_{AB}^{VV})^{1/2} \alpha_I \, .
 \label{4three}
\end{equation}
The main point is, now, that observable facts give not only very
stringent limits on $\eta_{AB}^{VV}$, but also mild limits on
$\alpha_I$. Indeed, Damour, Gibbons and Gundlach \cite{DGG,DG}
have shown, in the case of different scalar couplings to
visible and invisible matter, that cosmological data were putting
the limit $\alpha_I \equiv \sqrt{2} \beta_I < 0.71$. [See also the
later work \cite{FG} which considered only the case $\alpha_V \equiv
0$, $\alpha_I \not = 0$.] An even more stringent limit comes from
gravitational lenses. Indeed, in some gravitational lenses one
measures three different ``gravitational masses'': a ``virial'' mass
$G_* (1+\alpha_I^2) M$ linked to binary interactions, the
gravitational mass probed by the $X$-ray-emitting gas $G_*
(1+\alpha_V \alpha_I) M$, and the lensing mass $G_* M$ (light being
uncoupled to the new field, be it scalar or vector). The
coincidence, within better than 30~\%, of these three masses in some
systems \cite{Dar} gives the limit $\alpha_I < (0.30)^{1/2} =0.55$.
[Modulo the sign change $\alpha_I^2 \rightarrow -\alpha_I^2$, this
argument applies to the vector case and therefore improves upon the
limit $\alpha_I^{({\rm vector})} <1$ used in \cite{SAHS93}.] We can
apply the above limits to the case of the elemental compositions of
the Moon (silica) and the Earth (iron core plus silica mantle). A
laboratory approximation of this case (Si/Al versus Cu) has given
$\eta_{AB}^{VV} = (5 \pm 7) \times 10^{-12}$ \cite{Su94} so that, at
the one sigma level, $| \eta_{AB}^{VV}|^{1/2} < 3.4 \times 10^{-6}$.
Finally, we get for the maximum Moon-Earth ($1-2$, keeping our
previous labels) differential acceleration caused by a possible
anomalous coupling to dark matter (taking into account a further
factor $m_{\rm core} / m_{\rm Earth} =0.32$)
\begin{equation}
 \left( {\Delta a \over a} \right)_{12}^{VI} < 6.0 \times 10^{-7}
 \sqrt{|f_{AB}|} \, .
 \label{4four}
\end{equation}
Moreover, the fractional composition-dependence $f_{AB}$ (where
$A=Si O_2$, $B=Fe$) is generically expected to be small compared
to one\footnote{If $f_{AB}$ were larger than one, one should modify
our analysis above, and define more carefully the average value
$\alpha_V$.}. For instance, in dilaton models \cite{DP94} one finds
$f_{AB} \simeq 1.89 \times 10^{-5} [(E/M)_A - (E/M)_B]$ where
$E=Z(Z-1)/(N+Z)^{1/3}$. Here: $Z =$ atomic number, $N =$ neutron
number, $M =$ mass in atomic mass units. The only (physically
motivated) case, we know of, where $\alpha_A$ would exhibit a
significant fractional variation over the periodic table is the case
of appreciable coupling to lepton number $L$, or to $B-L=N$. [In
view of the small variation of the baryon to mass ratio $(B/M)_{AB}
\sim 10^{-3}$, any coupling involving $L$ with a relative
coefficient of order unity leads to essentially the same results.]
For instance, for a coupling to $B-L$, one gets $f_{AB} =
(2N/M)_{Si O_2} - (2N/M)_{Fe} \simeq -0.076$. Inserting this figure
in Eq.~(\ref{4four}) and using the full galactic acceleration
$a\simeq 1.9 \times 10^{-8}\; {\rm cm/s}^{2}$, one gets
\begin{equation}
 (\Delta a)_{12}^{VI} < 3.1 \times 10^{-15} {\rm cm/s}^{2} \, ,
 \label{4five}
\end{equation}
which is ten times smaller than the upper limit found in a recent
analysis of LLR data \cite{NMS95}. Even if we take $|f_{AB}| \sim
1$, we get $(\Delta a)_{12}^{VI} \alt 1\times 10^{-14}\; {\rm
cm/s}^{2}$, which is three times smaller than the result of
\cite{NMS95}. We conclude that, within what we consider the most
natural theoretical framework LLR data (and {\it a fortiori}
laboratory experiments \cite{Su94}) do not (yet) probe a
theoretically very significant domain of values of possible
anomalous couplings to dark matter.

Denoting ${\bf N}_G$ the projection of the unit vector directed
toward the galactic center on the ecliptic plane and ${\bf R}\equiv
(X,Y)$, the galactic polarization effect is described by the
potential
\begin{equation}
 R_G = A_G ({\bf N}_G.{\bf R}) \label{4six}
\end{equation}
analogous to Eqs.~(\ref{2four}). The parameter $A_G$
phenomenologically represents a differential acceleration of the
Moon and the Earth toward the galactic center. The corresponding
generating function $G$ is given by
\begin{equation}
 G = {\hat \omega} \left[\left(1+w\right)\zeta^{1+m}
  e^{-i\phi_G} + \left(1+{\bar w}\right)\zeta^{-(1+m)}
  e^{i\phi_G}\right]\; , \label{4seven}
\end{equation}
where
\begin{equation}
 {\hat \omega} \equiv m^2 {A_G \over {\tilde a} n'^2}\; .
 \label{4height}
\end{equation}
The polar angle $\phi_G =1.1^0$ gives the angular distance of
the galactic center from the lunar (and solar) position
corresponding to the above chosen new-moon phase at MJD51~521.2.
Employing Eq.~(\ref{3nine}) we obtain the source term of Hill's
problem in the following form
\begin{equation}
 S = -{\hat \omega} e^{i \phi_G}\zeta^{-(1+ m)} \; . \label{4nine}
\end{equation}
Because of the similarity of this function with (\ref{3eighteen}) we
recover the qualitative conclusions of Sec.~III.B. The sidereal
perturbation of the lunar orbit reads
\begin{equation}
 {\delta_\omega r\over {\tilde a}} = -2{{\hat \omega}\over m^2}
 S_{\rm gal}(m)\cos\left[n\left(t-t_0\right)- \phi_G\right] \; ,
 \label{4ten}
\end{equation}
with
\begin{eqnarray}
 S_{\rm gal}\left(m\right) &=& 1 -{75\over 8}m +{235\over 4}m^2
   - {127637\over 384}m^3 \nonumber \\
 & & + {4172299\over 2304}m^4 + {\cal O}(m^5) \; . \label{4eleven}
\end{eqnarray}
A more complete set of the coefficients of this series is given in
Table~\ref{tab4}. The dominant $m$-dependence of this series is
again captured by factorizing $[{\dot \varpi} (m)]^{-1}$. The
numerical value of the series (\ref{4eleven}) for the lunar orbit is
$S_{\rm gal}(m_{\rm Moon}) = 0.5050$. Clearly,
the dark matter differential coupling contributes also to the
$(n-2n')$-frequency of the radial oscillation of the lunar orbit. We
do not give here the detailed result, just quoting that its amplitude
is about $5.94$ times smaller than the amplitude of the principal
galactic polarization contribution (\ref{4ten}). Finally, the series
$S_{\rm gal}(m)$ shows the same pole, near $m_{cr} \simeq -0.18407$
as the sidereal series in (\ref{3nineteen}).

\section{ Conclusions}
The main results of this paper may be summarized as follows:
\begin{itemize}
\item We have confirmed, by more detailed computations, previous
 suggestions \cite{N73,DEF94} that LLR data have the potential of
 constraining the post-Newtonian parameter $\alpha_1$ at the $1\times
 10^{-4}$ level or better. We showed that the preferred frame
 perturbations associated with the $\alpha_1$ parameter contribute
 a large spectrum of frequencies in the radial oscillation of
 the lunar orbit. The dominant $\alpha_1$-effects occur at
 frequencies $n$ (sidereal effect) and $n'$ (yearly effect) with
 well-determined phases, and there is a sub-dominant effect at
 frequency $n-2n'$. Although the analytical results that we
 obtained from a high-order Hill-Brown algorithm should be accurate
 enough for fitting purposes, it may be advisable to resort to a
 direct numerical integration of the equations of motion (see e.g.
 \cite{DEF94} for the $\alpha_1$ contributions to the equations of
 motion).

\item We found that retrograde planar orbits with $n'/(n-n')
 =-0.18407$ (which have fixed perigees in inertial space) exhibit a
 resonant amplification of preferred-frame effects. Putting an
 artificial satellite near such an orbit could be an efficient
 (though expensive) way of improving the present bounds on $\alpha_1$.

\item We have extended our analysis to another
 perturbation linked to a fixed direction in space: namely, a
 possible differential acceleration toward the galactic center.
 Evidently, this perturbation exhibits also a pole at $m_{cr}
 =-0.18407$ corresponding to an orbit ``frozen in space''. We argue,
 however, that there are strong {\it a priori} theoretical
 constraints on the conceivable magnitude of such an effect.
\end{itemize}

A specific suggestion for future work concerns applying our analysis
of the singularly perturbed spatially frozen orbits to
planetary satellites. It is widely known that the solar system
satellites are submitted to a complicated cosmogonic tidal
evolution. It might be interesting to study if some of these
bodies evolved historically through such a frozen orbit
configuration yielding indirect limits on the $\alpha_1$ parameter.
For instance, one easily verifies that several of Jupiter small
satellites do lie close to the frozen configuration. A careful
study of these problems is, however, beyond the scope of this
paper.

\acknowledgments
We thank M. Moons for providing us with the algebraic manipulator
MINIMS. C. Veillet is thanked for informative discussions. We are
grateful to J.G.~Williams for pointing out that the annual
perturbation might be well separated, and to J.~M\"uller for
detecting an error in our computation of the phases $\phi$ and
$\phi_G$. D.V.
worked on this paper while staying at the OCA/CERGA,
Grasse (France) and being supported by an H. Poincar\'e research
fellowship. He is also grateful to IHES, Bures sur Yvette (France)
for its kind hospitality and partial support.

%%%%%%%%%%%%%%%%%%%%%%%%%%%%%%%%%%%%%%%%%%%%%%%%%%%%%%%%%%%%%
%
%          TABLES:
%
%%%%%%%%%%%%%%%%%%%%%%%%%%%%%%%%%%%%%%%%%%%%%%%%%%%%%%%%%%%%%

\begin{table}
\caption{ Coefficients $s_k$ of the $S^{(1)}_{\alpha_1}(m)$
 series in powers of $m$. The percentage $p_k$ of the
 contribution of the listed terms to the series
 for the lunar orbit -- $m=m_{\rm Moon}=0.0808489375\ldots$ --
 is given in the second column. The last column gives the
 ratio $(s_{k-1}/s_k)$.}
\begin{tabular}{rccc}
 $k$ & $s_k$ & $p_k$ & $(s_{k-1}/s_k)$ \\
\tableline
0 &  1.000000000000000 &  --     & --           \\
1 & -8.375000000000000 & -67.71  & -0.119402985 \\
2 &  49.37500000000000 &  32.27  & -0.169620253 \\
3 & -268.2473958333333 & -14.18  & -0.184065161 \\
4 &  1444.161892361111 &   6.17  & -0.185746070 \\
5 & -7841.277434172453 &  -2.71  & -0.184174314 \\
6 &  42586.57408613037 &   1.19  & -0.184125575 \\
7 & -231375.2814872327 &  -0.52  & -0.184058443 \\
8 &  1256984.833107015 &   0.23  & -0.184071657 \\
9 & -6828885.820821555 &  -0.10  & -0.184068802 \\
\end{tabular}
 \label{tab1}
\end{table}

\begin{table}
\caption{ Coefficients $s^\prime_k$ of the
 $S^{\prime(1)}_{\alpha_1}(m)$ series in powers of $m$.
 Other parameters as in Table~I.}
\begin{tabular}{rccc}
 $k$ & $s_k$ & $p_k$ & $(s_{k-1}/s_k)$ \\
\tableline
0 &  1.000000000000000 &  --    & --           \\
1 & -7.166666666666667 & -57.94 & -0.139534884 \\
2 &  40.09305555555556 &  26.21 & -0.178750823 \\
3 & -216.8476851851852 & -11.46 & -0.184890401 \\
4 &  1173.979870756172 &   5.02 & -0.184711587 \\
5 & -6371.734190136318 &  -2.20 & -0.184248093 \\
6 &  34622.85594255030 &   0.97 & -0.184032600 \\
7 & -188086.0224857120 &  -0.42 & -0.184079899 \\
8 &  1021827.835270500 &   0.19 & -0.184068212 \\
\end{tabular}
 \label{tab2}
\end{table}

\begin{table}
\caption{ The same as in Table~I but for the $S^{(3)}_{\alpha_1}(m)$
 series (ratio of the consecutive coefficients omitted).}
\begin{tabular}{rcc}
 $k$ & $s_k$ & $p_k$ \\
\tableline
0  &  1.000000000000000 &  --    \\
1  & -2.000000000000000 & -16.17 \\
2  & -1.953125000000000 &  -1.28 \\
3  & -5.875000000000000 &  -0.31 \\
4  &  12.50295003255209 &   0.05 \\
5  &  49.86885579427062 &   0.02 \\
6  &  73.34703290903986 &   $<$  \\
7  & -113.6395324400916 &   $<$  \\
8  & -670.4626529838962 &   $<$  \\
9  & -1047.599614057912 &   $<$  \\
\end{tabular}
 \label{tab3}
\end{table}

\begin{table}
\caption{ The same as in Table~I but for the $S_{\rm gal}(m)$
 series.}
\begin{tabular}{rccc}
 $k$ & $s_k$ & $p_k$ & $(s_{k-1}/s_k)$ \\
\tableline
0 &  1.000000000000000 & --     & --           \\
1 & -9.375000000000000 & -75.80 & -0.106666667 \\
2 &  58.75000000000000 &  38.40 & -0.159574468 \\
3 & -332.3880208333333 & -17.57 & -0.176751256 \\
4 &  1810.893663194444 &   7.74 & -0.183549166 \\
5 & -9846.043167679392 &  -3.40 & -0.183920955 \\
6 &  53488.73992091048 &   1.49 & -0.184076932 \\
7 & -290602.0005454238 &  -0.66 & -0.184061843 \\
8 &  1578765.334003927 &   0.29 & -0.184069155 \\
9 & -8577020.610670500 &  -0.13 & -0.184069201 \\
\end{tabular}
 \label{tab4}
\end{table}

%%%%%%%%%%%%%%%%%%%%%%%%%%%%%%%%%%%%%%%%%%%%%%%%%%%%%%%%%%%%%
%
%          FIGURES:
%
%%%%%%%%%%%%%%%%%%%%%%%%%%%%%%%%%%%%%%%%%%%%%%%%%%%%%%%%%%%%%

\begin{figure}
\caption{ Amplitude (in centimeters) $C_{\alpha_1}$ of the sidereal
 oscillation (for $\alpha_1 = 1$) vs. the Hill parameter $m$ (positive
 for prograde orbits, negative for retrograde orbits). For high
 orbits $C_{\alpha_1} \simeq \case23 {\tilde a}{\hat \epsilon}_1
 S^{(1)}_{\alpha_1}(m) m^{-2}$ as given in Eq.~(3.19). In the case
 of low orbits we introduce the influence of the Earth multipolar
 structure by adding a quadrupole contribution to the denominator
 of a Pad\'e approximant of $S^{(1)}_{\alpha_1}(m)$. Besides the
 singularly amplified orbit at $m_{cr} = -0.18407$, two celestial
 bodies are indicated: (i) the Moon $(M)$, and (ii) the LAGEOS
 satellite $(L)$.
 \label{fig1}}
\vspace{6cm}
%\begin{center}
%\begin{picture}(100,100)(145,-682)
%\special{em:graph dv3.pcx}
%\end{picture}
%\end{center}
\end{figure}

\end{document}